\documentclass[final,12pt, twocolumn]{elsarticle}

\usepackage{bm}

\usepackage{amsmath}
\usepackage{color}
\usepackage{textcomp}
 \usepackage{graphicx}

\usepackage{amssymb}

\usepackage{todonotes}

\journal{Ultramicroscopy}

\begin{document}





\providecommand{\lap}{\ensuremath{\nabla^2}}
\providecommand{\vb}{\ensuremath{V_{\text{bias}}}} 
\providecommand{\vr}{\ensuremath{V(\mathbf{r})}}
\providecommand{\ve}{\ensuremath{V_0}}
\providecommand{\vs}{\ensuremath{V}}
\providecommand{\Ee}{\ensuremath{E_{0}}}
\providecommand{\er}{\ensuremath{E_0(\mathbf{r})}}
\providecommand{\ern}{\ensuremath{E_0(\mathbf{r}_0)}}
\providecommand{\der}{\ensuremath{\Delta E_0(\mathbf{r})}}
\providecommand{\vemr}{\ensuremath{V_{\text{MMT}}}}
\providecommand{\Ur}{\ensuremath{U(\mathbf{r})}}
\providecommand{\wf}{\ensuremath{\Delta \Phi (\mathbf{r})}}
\providecommand{\wfs}{\ensuremath{\Delta \Phi_{\text{g-s}} (\mathbf{r})}}
\providecommand{\um}{\,\textmu m}
\providecommand{\rn}{\ensuremath{\mathbf{r}_0}}

\title{Low-Energy Electron Potentiometry}

\date{\today}

\author[leiden,columbia]{Johannes Jobst$^\star$}
\author[leiden]{Jaap Kautz$^\star$ }
\author[leiden]{Maria Mytiliniou}
\author[leiden,ibm]{Rudolf M.\ Tromp}
\author[leiden]{Sense Jan van der Molen}
\address[leiden]{Huygens-Kamerlingh Onnes Laboratorium, Leiden University, P.O. Box 9504, NL-2300 RA Leiden, Netherlands}
\address[columbia]{Department of Physics, Columbia University, New York, New York 10027, USA}
\address[ibm]{IBM T.J. Watson Research Center, 1101 Kitchawan Road, P.O. Box 218, Yorktown Heights, New York 10598, USA}

\begin{abstract}
In a lot of systems, charge transport is governed by local features rather than being a global property as suggested by extracting a single resistance value.
Consequently, techniques that resolve local structure in the electronic potential are crucial for a detailed understanding of electronic transport in realistic devices.
Recently, we have introduced a new potentiometry method based on low-energy electron microscopy (LEEM) that utilizes characteristic features in the reflectivity spectra of layered materials \cite{Kautz-LEEP}. Performing potentiometry experiments in LEEM has the advantage of being fast, offering a large field of view and the option to zoom in and out easily, and of being non-invasive compared to scanning-probe methods.  However, not all materials show clear features in their reflectivity spectra. 
Here we, therefore, focus on a different version of low-energy electron potentiometry (LEEP) that uses the mirror mode transition, i.e. the drop in electron reflectivity around zero electron landing energy when they start to interact with the sample rather than being reflected in front of it. 
This transition is universal and sensitive to the local electrostatic surface potential (either workfunction or applied potential). It can consequently be used to perform LEEP experiments on a broader range of material compared to the method described in Ref. \cite{Kautz-LEEP}. 
We provide a detailed description of the experimental setup and demonstrate LEEP on workfunction-related intrinsic potential variations on the Si(111) surface and for a metal-semiconductor-metal junction with external bias applied.  
In the latter, we visualize the Schottky effect at the metal-semiconductor interface.
Finally, we compare how robust the two LEEP techniques discussed above are against image distortions due to sample inhomogeneities or contamination. 

\end{abstract}



\maketitle

\renewcommand{\thefootnote}{$\star$} 
\footnotetext{These authors contributed equally.}

\section{Introduction}\label{sec:introduction}	
While conductance measurements are commonplace in material characterization, such measurements are typically limited to averages over entire samples whereas local variations of conductance are lost.
This poses limitations in the analysis of local effects which often govern global transport. For example, in topological insulators the current is completely carried by edge channels. Moreover, random charge inhomogeneity causes lateral variations in the doping level of semiconducting materials and thin films while step edges and grain or domain boundaries can locally influence their conductivity.
This is particularly important for two-dimensional electron systems that are easily perturbed due to their large surface/bulk ratio. With the emergence of a novel material class of atomically layered crystals such as graphene, this is more relevant than ever.
A good understanding of transport properties therefore requires a technique which can locally probe electrical conductance.

Several tools can be used to study charge transport by locally probing the electrical potential at the surface. 
Photoelectron emission microscopy (PEEM) \cite{Ballarotto-PEEM-dopants, Giesen-PEEM-pn} and synchrotron-based scanning PEEM \cite{ Phaneuf-SPEM-pn}, for example, were used to study doping levels and band bending in semiconductor p-n junctions. 
And while scanning probe techniques such as Kelvin probe force microscopy \cite{ tzalenchuk-kelvin-probe}, scanning squid microscopy \cite{nowack-imaging-SHE} and scanning tunneling microscopy \cite{Park-STM-pn, ross-steps} have provided exciting new results, their acquisition times are long and their field of view is limited due to the scanning nature of the methods.
In the faster scanning electron microscopy, potentiometry measurements can be performed by studying the effect of the local sample potential on the secondary-electron trajectories \cite{Oatley1957,Menzel1983,Girard1988}. 
However, the secondary-electron emission is strongly material dependent \cite{Nakamae1981,Girard1988} and the high landing energy (0.5--10\,keV) of the electrons can severely influence the local conductivity\cite{Nakamae1981,Girard1988}, making the results difficult to interpret. 
This problem can be overcome in low-energy electron microscopy (LEEM) where the electron landing energy is typically in the range of 0--30\,eV. LEEM-based potentiometry (LEEP) thus provides a fast alternative with a large field of view. 
Anderson and Kellogg used LEEM to probe potential differences in doped silicon samples \cite{Kellogg-pn}, and we showed how the sensitivity of LEEP to the local potential can be further enhanced \cite{Kautz-LEEP} by studying the full energy dependence of electron reflection.

In the present paper we describe a generalized version of LEEP, which uses the LEEM Mirror-Mode Transition (MMT) -- a steep drop in intensity as we change the sample potential relative to the electron gun potential from a more negative value where all electrons are reflected above the sample, to a more positive value where they interact with the surface. 
We therefore, call this method mirror-mode LEEP (M-LEEP). The MMT is present for all materials while clear LEED (Low Energy Electron Diffraction) IV-curves as used in Ref. \cite{Kautz-LEEP} are not generally available. Consequently, M-LEEP can be applied to a broader range of materials, of which we show two applications in this paper. 
First, we demonstrate M-LEEP by measuring the intrinsic potential differences between different surface reconstructions of the Si(111) surface \cite{Hannon-phase-transition}. 
Second, we locally probe the potential distribution within a metal-semiconductor-metal junction over which an external bias is applied.
This allows us to directly probe the Schottky effect without the use of metallic probe contacts, which could easily disturb such a measurement.

\begin{figure}[t]
	\includegraphics[width=\columnwidth]{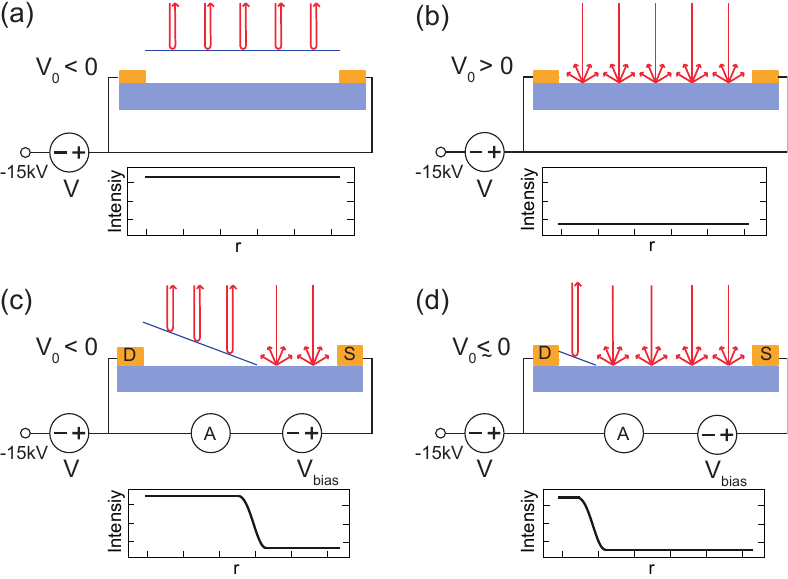}
	\caption{
		(a)	For negative voltages \ve{}, all electrons are reflected and the reflection intensity is high for all positions.
		(b)	For positive \ve{}, electrons scatter off the surface reducing the reflection intensity.
		(c)	When a bias voltage \vb{} is applied between source and drain, the landing energy and the reflection intensity become position dependent. 
		At the MMT, i.e., where the landing energy is zero, the intensity drops sharply.
		(d) By varying \ve, the position of the MMT shifts.
		By determining the value of \ve\ of the MMT for each position $r$, the local potential \vr{} can be determined.
		}
	\label{fig:mleep}
\end{figure}
	
\section{Potentiometry in LEEM}\label{sec:potentiometryinleem}
In LEEM, the sample is illuminated with a parallel beam of low-energy electrons. The electrons are extracted from the gun and accelerated to 15\,keV. While they travel through most of the electron-optical system at this high energy, their low kinetic landing energy $\Ee \approx 0$--30\,eV is achieved by decelerating them towards the sample. For that purpose, the sample is lifted to $-15\text{\,kV} + \vs$. By changing the voltage \vs{}, the landing energy \Ee{} of the electrons can thus be accurately tuned.
A bright-field LEEM image is formed by the cathode objective lens, projecting the specularly reflected electrons with energy \Ee{} onto a channelplate-intensified detector.

The reflectivity, and therefore, the image intensity $I$, in LEEM experiments is a strong function of \Ee{}. 
The intensity of the (0,0) LEED beam with energy (the so-called LEED-IV curve) is highly sensitive to surface atomic structure \cite{hannon-LEEM-IV, Flege-IV}, as well as the unoccupied band structure \cite{Jobst-ARRES, Jobst-ARRES-GonBN} of the sample under investigation.
One feature present in the IV-curves for all materials is the MMT when the landing energy is changed from negative to positive values.
In mirror mode (MM), i.e.\ for negative \Ee{}, all incoming electrons are repelled and reflected before they reach the sample (Fig.\ \ref{fig:mleep}a). This leads to a high intensity on the detector. 
For positive \Ee{}, the electrons are decelerated less strongly and reflect from the sample surface (Fig.\ \ref{fig:mleep}b) causing a reduced intensity on the detector.
The MMT is therefore characterized by a steep drop in specular reflection intensity that can easily be recognized in the IV-curve.
This intensity drop can be described by an error function (erf), if you assume a Gaussian energy distribution of the incident electrons, the center of which defines the voltage \vemr{} for which the average landing energy is zero.

During regular LEEM imaging of a homogeneous sample, the landing energy and therefore also \vemr{} is the same for the whole illuminated area. In a typical LEEP experiment, in contrast, the surface potential \vr{} is inhomogeneous over the sample and therefore, the local landing energy \er{} becomes a function of position 
\begin{equation}
\er = \ve - e \left[\wf + \Ur\right],
\end{equation} 
where \wf{} is the spatial variation of the workfunction and \Ur{} the spatial variation of the surface potential due to an applied bias voltage \vb{} between $\mathbf{r}$ and a reference point \rn{} (e.g.\ in Sec.\ \ref{sec:WF}, a point on the $7\times 7$-reconstruction).
The decelerating voltage \ve{} is defined such that $\ve := 0$ when $\ern = 0$.
A position dependence of \wf{} can arise due to workfunction differences within the surface, e.g.\ by the coexistence of different surface reconstructions. While \wf{} is an intrinsic property of the studied surface, \Ur{} is defined by how \vb{} drops over the sample (Fig.\ \ref{fig:mleep}c) and therefore, contains information about local conductance. 
Both these properties can be investigated by measuring the landing energy locally resolved. 

In M-LEEP we measure \er{} by acquiring LEEM images while scanning the voltage \ve{} and using the steep intensity drop at the MMT to determine \vemr{} for each position. The difference in \vemr{} is then exactly the difference in local landing energy and hence local surface potential \vr{} between those two points. As an example, we sketch in Fig.\ \ref{fig:mleep}c and d how the position where $\er = 0$ shifts over the sample for different values of \ve{}, when an external bias is applied.

In this article, we discuss the two fundamental cases. First we apply no external bias $\Ur = 0$ and study workfunction variations within the sample. Second, we apply an external bias over a sample with no lateral workfunction variations $\wf = 0$. For the first case, we study the silicon (111) surface that shows different reconstructions with different surface potentials. For the second case, we look at the voltage drop between metallic electrodes on unreconstructed silicon.

\begin{figure}[t]
	\includegraphics[width=\columnwidth]{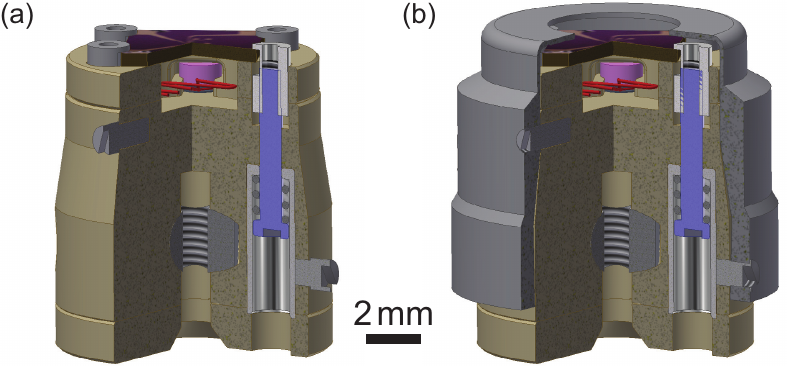}
	\caption{
		(a)
		Construction drawing of the LEEP puck with four pins making electrical contact to the sample corners. They are pulled down by spring-loaded screws (blue) to maintain good contact during heating experiments. One of them is connected to the cap with a screw to serve as local potential reference.
		(b)
		A standard SPECS molybdenum cap \cite{Tromp-stage} that shields sharp corners at the sample and the contacts is fixed on the puck by screws such that a 200\um\ wide gap isolates it from the contact pins. A tungsten filament (red) below the sample serves as electron bombardment heater.
	}
	\label{fig:puck}
\end{figure}
  
\section{LEEP sample holder and electronics}
\label{sec:design}
Our experiments are performed at the ESCHER facility \cite{ESCHER}, which is based on the commercially available, aberration-corrected FE-LEEM P90 instrument (SPECS GmbH, Berlin) designed by IBM \cite{tromp-AC1}. In order to perform a LEEP experiment, a number of additional requirements compared to standard LEEM have to be fulfilled and the experimental setup has to be adjusted accordingly. 
First, an in-plane bias voltage has to be applied across the sample while imaging.  
Second, the current through the sample needs to be measured to be able to compute the local conductivity from the derived surface potential.  
Third, 4-probe voltage measurements have to be performed \emph{in situ} to compare the LEEP results to the global conductance.
For that purpose, we developed special measurement electronics and a new puck that holds the sample and makes electrical contact to it. The puck is milled from aluminum oxide and thus, is non-magnetic and compatible with ultra-high-vacuum (UHV). It contains an electron bombardment heater for temperatures up to 1450\,K, fits onto the existing SPECS sample holder (described in Ref.\ \cite{Tromp-stage}) and is only $\approx\,1\,\text{cm}^3$ small (cf.\ Fig.\ \ref{fig:puck}a).
A bayonet fitting at the side of the cap allows for \emph{in situ} mounting of the puck on the macor carrier piece \cite{Tromp-stage}. 
	
The sample is mounted onto the puck and is electrically contacted by molybdenum pins, which are pressed onto the corners of the sample by springs (Fig. \ref{fig:puck}b). Metallic contact pads in the corners of the sample lead towards its center where the devices are structured and LEEP microscopy is performed.  
The contact pins are concealed below the standard SPECS sample cap to prevent field enhancement at their edges that could cause arcs. 
The cap is lifted to $\approx -15$\,kV where the precise voltage between cap and objective lens controls the landing energy of the electrons in LEEM/LEEP (see Sec.\ \ref{sec:potentiometryinleem}). In order to make the cap potential the local reference potential (as in standard LEEM), the cap is fixed onto the puck such that it is electrically connected to one of the contact pins. It is insulated from the other contacts by a 200\um\ wide gap (cf.\ Fig.\ \ref{fig:puck}c).
Apart from the four connections to the sample, two more contacts lead to the tungsten filament of the electron bombardment heater. 

Under operating conditions of the microscope, the sample is kept at a potential of $\approx -15$\,kV and thus cannot be directly connected to a conventional power supply to apply \vb. Consequently, the electronics are divided in a low-voltage and a high-voltage part that are electronically decoupled through fiber-optical connections. 
The applied voltage as well as the measured voltages and currents are optically transmitted by a frequency modulated signal via these fibers. All electronic components that are in direct contact with the sample are equipped with an overvoltage protection consisting of a spark gap placed parallel to two Zener diodes. It diverts the current to the local ground in the case of arcing due to the strong electric field between sample and objective lens. 

\begin{figure}
	\includegraphics[width=\columnwidth]{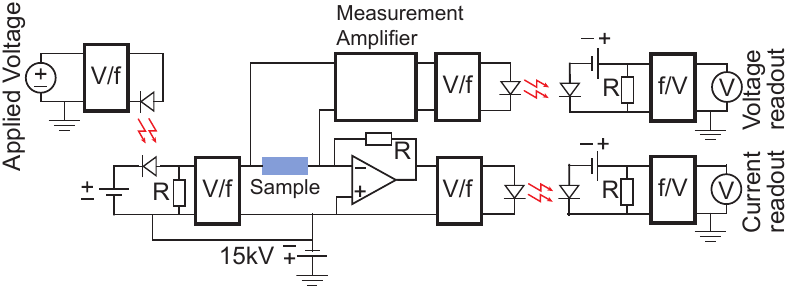}
	\caption{
		The electronics are divided in a high- and a low-voltage part, which are optically decoupled. The high-voltage part applies \vb\ over the sample and measures the current through the sample as well as the two-point and four-point voltage drop.
		}
	\label{fig:electronics}
\end{figure}

\section{Results}\label{sec:results}
\subsection{Intrinsic Potential Differences of the Si(111) Surface\label{sec:WF}}
We use the Si(111) surface that exhibits two surface structures with intrinsic surface potential differences \cite{Hannon2003} as a test case to verify the potential resolving capabilities of LEEM. 
Above 1135\,K, Si(111) shows the unreconstructed 1$\times$1 phase, while a 7$\times$7 reconstruction is formed when the temperature is lowered.
Around the phase transition temperature, the surface forms a mixture of 7$\times$7 and 1$\times$1 reconstructed domains \cite{Hannon2003}. 
We prepare these mixed surfaces by flash annealing Si(111) samples several times to $\approx 1450$\,K for 10 seconds to remove the native oxide and then cool down to the desired temperature.	

\begin{figure}[ht]
		\includegraphics[width=\columnwidth]{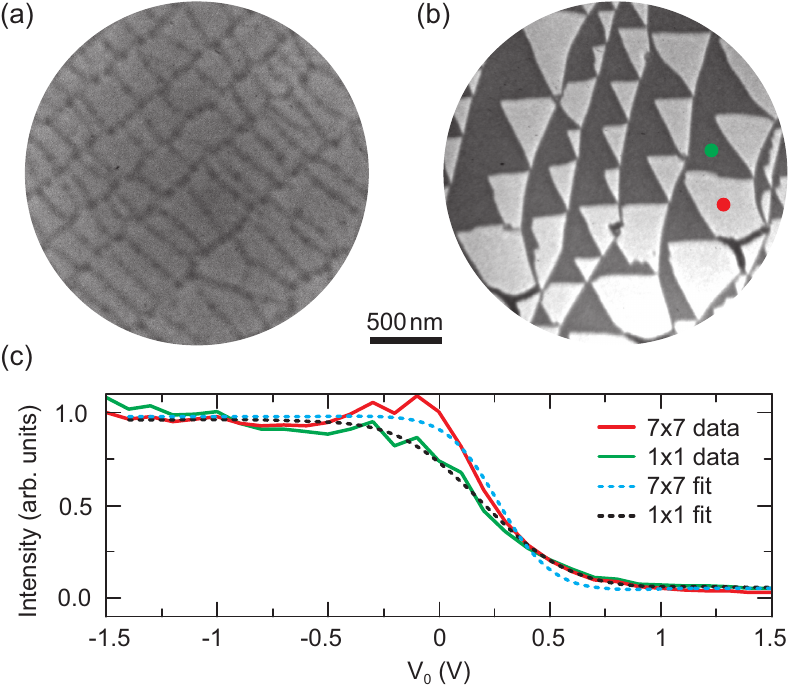}
		\caption{
		(a) Bright-field LEEM micrograph ($\Ee = 1.6$\,eV) showing the almost completely 7$\times$7 reconstructed surface (bright) at 870\,K. Small 1$\times$1 areas exist only at step edges (dark lines from bottom left to top right) and domain boundaries (dark lines perpendicular to step edges).
	(b) At 1040\,K a mixture of triangular 7$\times$7 domains (bright) and 1$\times$1 reconstructed areas (dark) are observed ($\Ee = 2.0$\,eV). 
	(c) IV-curves taken from two different areas marked in (b) show a clear MMT (solid red and green for 7$\times$7 and 1$\times$1 reconstruction, respectively). An error function (dashed) is fitted to the data to determine \vemr.}
		\label{fig:mixedmm}
		\end{figure}	

Figure \ref{fig:mixedmm}a shows a real-space bright-field image of a Si(111) surface at 870\,K. The surface is widely 7$\times$7 reconstructed with 1$\times$1 domains only at the step edges. Domain boundaries are visible as dark lines perpendicular to the step edges. 
At 1040\,K, around the phase transition temperature, the surface exhibits triangular 7$\times$7 domains surrounded by 1$\times$1 reconstructed areas (Fig.\ \ref{fig:mixedmm}b).
		
\begin{figure}[t!]
		\includegraphics[width=\columnwidth]{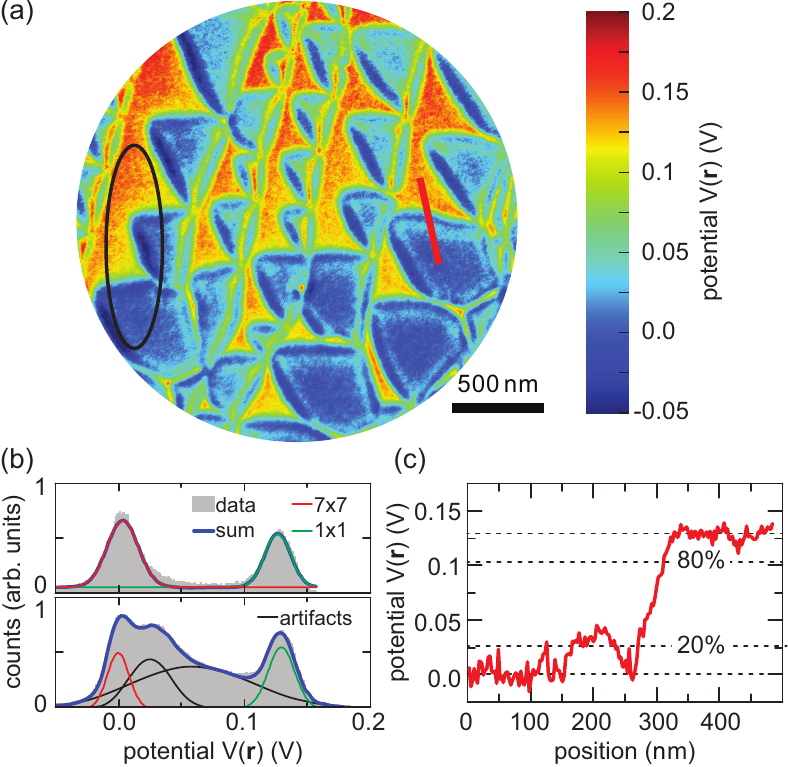}
		\caption{
	(a) A potential map of the area in Fig.\ \ref{fig:mixedmm} clearly resolves that 7$\times$7 reconstructed domains have a higher surface potential than the surrounding 1$\times$1 areas. 
	(b) Histograms of the potential values in (a) shows that this difference is $0.12 \pm 0.02$\,V. This is more clear from the top histogram gathered from the area marked black in (a), whereas two more peaks (black lines) are visible in the bottom histogram that is taken over the full field of view. They arise from artifacts due to lateral fields at domain boundaries (visible as bright blue and green lines in (a)).
	(c) From a line scan of the area indicated in (a), the spatial resolution, defined as the lateral distance between 20\% and 80\% of the step height, is determined to be 40\,nm. The determined workfunction difference is 0.13\,eV. 
		}
		\label{fig:potmap}
		\end{figure}

We measure the difference in the surface potential of these surface reconstructions, by varying the sample potential via \ve{} while imaging the sample as described in Sec.\ \ref{sec:potentiometryinleem}.
In Fig.\ \ref{fig:mixedmm}c, the IV-curves for the positions marked in Fig.\ \ref{fig:mixedmm}b clearly show the MMT as a strong drop in intensity.
The fact that this transition is not infinitely sharp, is caused by the energy spread of the electrons leaving the gun. 
If we assume this spread to be Gaussian, we can deduce the standard deviation of the electron energy distribution as 0.15\,eV from the width of the fitted error function (dashed lines in Fig.\ \ref{fig:mixedmm}c).
The center of the fitted error function defines \vemr{} for which the local landing energy $\er = 0$.
Repeating this procedure for each pixel individually, we obtain a map of the surface potential \vr{} of the sample as shown in Fig.\ \ref{fig:potmap}a. 
Since no external bias is applied over the sample, the local variations are caused by workfunction differences between the different surface reconstructions. 
The domains of unreconstructed and 7$\times$7-reconstructed that are visible in Fig.\ \ref{fig:mixedmm}b, clearly show different workfunction values in this LEEP image. In the vicinity of step edges and impurities, however, artifacts caused by lateral field inhomogeneity is observed as well.
		
In order to quantify the workfunction difference between different domains, we plot histograms of the local potential values for this sample in Fig.\ \ref{fig:potmap}b.
The top histogram is taken from the area indicated in black in Fig.\ \ref{fig:potmap}a that is free from field-induced artifacts. It shows one peak from the $7 \times 7$ reconstruction at lower surface potential and a second one from the $1 \times 1$ areas, clearly separated from one another.
From this splitting, we determine the difference in surface potential to be $0.12 \pm 0.02$\,V, which is comparable with the $0.15 \pm 0.03$\,eV found by Hannon \emph{et al.} \cite{Hannon-phase-transition}. 
The bottom histogram in Fig.\ \ref{fig:potmap}b is generated from the full field of view. In addition to the 1$\times$1 and the 7$\times$7 peaks, two more peaks are needed to describe the histogram. Those values are caused by the abovementioned artifacts and can obscure the data evaluation easily.
The spatial resolution of M-LEEP is determined by looking at a line scan over the 7$\times$7-1$\times$1 interface (see Fig.\ \ref{fig:potmap}c). The height of the step is $0.13 \pm 0.02$\,eV in agreement with the evaluation from the histogram in Fig.\ \ref{fig:potmap}b and literature.
From the lateral spacing between 20\% and 80\% lines of the step in potential, a resolution of 40\,nm is extracted. 
While here, the LEEM images are slightly out of focus, we expect the spatial resolution of the potentiometry measurements to be closer to the optimal LEEM resolution ($< 2$\,nm) \cite{schramm-thesis} for optimal alignment conditions.
With M-LEEP, however, this optimal microscopy resolution cannot be reached since the M-LEEP resolution will always be limited by the presence of a lateral electric field above the 7$\times$7-1$\times$1 interface, induced by the difference in surface potential \cite{Kennedy-MM-distortions, Kennedy-MM-distortions2}. We estimate 10\,nm to be the resolution limit of M-LEEP.
This experiment nevertheless clearly demonstrates how M-LEEP can be used to measure local changes in the workfunction of a surface.

\subsection{Electron Transport Through a Metal-Semiconductor-Metal Junction \label{sec:Schottky}}
In order to study the potential distribution due to an external applied bias with M-LEEP, we prepare an undoped Si(111) sample ($\rho=10\,$k$\Omega$cm) with two metal contacts. 
Dust particles are removed from the sample with acetone and isopropanol.
Then two contacts separated by 4\,\textmu{}m are fabricated by shadow evaporation of 5\,nm of chromium and 30\,nm of Au using a tungsten wire as a mask, thereby minimizing contamination of the sample.
Subsequently, the sample is cleaned by nitric acid ($97\%$ Sigma Aldrich) for 10 seconds, rinsing in demineralized water and a 30 second treatment in buffered oxide etch [7:1 ammonium fluoride (40\% in water) and hydrofluoric acid (49\% in water), J.T.\ Baker]. The latter removes the native oxide and passivates the Si surface with hydrogen.
The sample is then directly loaded into the LEEM to prevent oxidation and contamination and heated to $\approx$\,600\,K \emph{in situ}.
		
\begin{figure}[t]
	\includegraphics[width=\columnwidth]{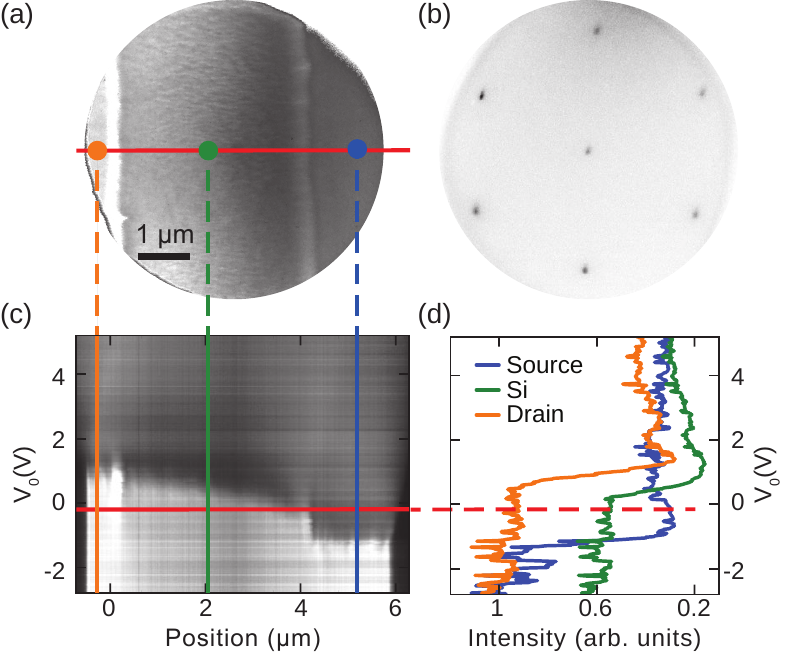}
	\caption{
	(a) LEEM image of the metal-semiconductor-metal junction at \ve{}=0.3\,eV and an applied bias of $\vb = 2$\,V. The center of the gap just goes through the MMT.
	(b) LEED pattern of the Si(111) surface within the junction. 
	(c)	Line scans at the line indicated in (a) for different values of \ve{} visualize the influence of the bias voltage. A linear voltage drop over the silicon and a large jump in voltage at the right metal-Si interface are apparent. 
	(d)	IV-curves taken at the positions indicated by circles in (a) corresponding to vertical cuts in (c). The voltage difference between the different points is visible as an energy shift of the MMT. 
	}
	\label{fig:slices}
\end{figure}

A LEEM image of the resulting two-probe structure is shown in Fig.\ \ref{fig:slices}a with the drain and source contacts on the left and right, respectively. The central Si(111) surface is clean and free of oxide as confirmed by the LEED pattern in Fig.\ \ref{fig:slices}b, proving the effectiveness of the cleaning method used. 		
Figure \ref{fig:slices}a is recorded at $\ve{} = 0.3$\,V with an external bias of $\vb{} = 2$\,V applied, resembling the situation in Fig.\ \ref{fig:mleep}c.
As a consequence, the left side of the sample appears much brighter than the right side because the former is in MM where all electrons are reflected (bright), while the latter is in imaging conditions where electrons scatter from the sample surface (darker). 
In between, the position of the MMT can be found, or in other words, a position where the local landing energy is zero. 
This MM-position shifts over the sample when \ve{} is changed as sketched in Fig.\ \ref{fig:mleep}c,d. 
The evolution of the MM-position is visualized in Fig.\ \ref{fig:slices}c where intensity profiles taken parallel to the current paths (along the red line in Fig.\ \ref{fig:slices}a) are plotted for different \ve. 
It is noteworthy that this is similar to plotting IV-curves for every position on the line.  Figure \ref{fig:slices}d, for example, shows IV-curves for the three points indicated by circles in Fig.\ \ref{fig:slices}a.
For low values of \ve{}, the whole sample is in MM and the intensity is high everywhere in (cf.\ left side of Fig.\ \ref{fig:slices}c), while for high values of \ve{} the intensity is lower throughout the device. 
The boundary between dark and bright is the MMT which is a direct visualization of the local potential $\vr{}$. 
Two features are evident in the lateral change of the local potential shown in Fig.\ \ref{fig:slices}c. First, a linear potential gradient over the Si is visible as expected for a homogeneous material. Second, the potential exhibits pronounced steps at the metal-Si interfaces.
		
\begin{figure}[t]
	\includegraphics[width=\columnwidth]{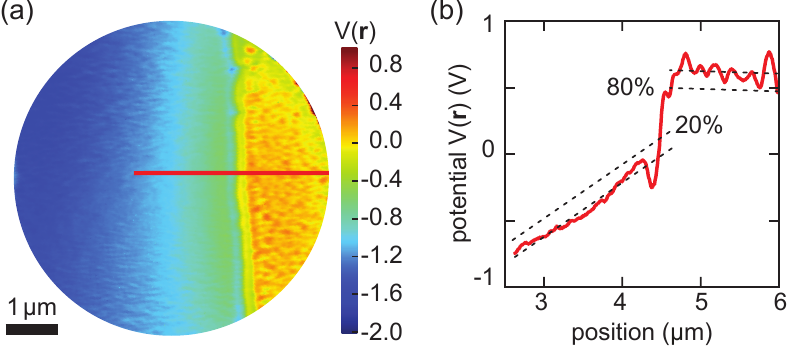}
	\caption{
	(a)
	Potential map of metal-Si interface at the biased source electrode.
	(b)
	Line scan along the line indicated in (a).
	From the errors in the fits to the line scan, a voltage resolution of 40\,mV and 60\,mV is derived for the Si and Au areas, respectively.
	The spatial resolution is determined to be 100\,nm from the lateral spacing between 20\% and 80\% of the step height. 
	}
	\label{fig:potmapgap}
\end{figure}

For a more rigorous, quantitative investigation, the automated determination of the MMT via IV-curves that is described in Sec.\ \ref{sec:WF} is used for every pixel in Fig.\ \ref{fig:slices}a to calculate a map of the local electrical potential (Fig.\ \ref{fig:potmapgap}a). Figure \ref{fig:potmapgap}b shows a line scan parallel to the current flow (along the red line in Fig.\ \ref{fig:potmapgap}a). It confirms the linear voltage drop over the silicon and the steep potential step at the metal-Si interface of the source contact that are already visible in Fig. \ref{fig:slices}c. 
From this step we estimate the lateral resolution of M-LEEP to be $\sim$100\,nm using the same 80\%-20\%-criterion as in Fig.\ \ref{fig:potmap}c.
Here, the lateral resolution is mainly limited by the large height difference of $\approx 35$\,nm between Si surface and Au electrode and a not perfectly clean sample. In addition, the lateral electrical field due to the difference in surface potential deforms the image.  
These effects are particularly pronounced for M-LEEP because electron trajectories are affected by asperity and impurities of the surface strongest around the MM as then, the electron energy is minimal \cite{Kennedy-MM-distortions, Kennedy-MM-distortions2}. 

We take the measured potential fluctuations around the expected trends as an estimate of the resolution in electrical potential. To obtain those, we fit linear trends to the data within the Au contact and the pure Si area, respectively. 
We take the standard deviation of the residuals of that fit as an upper bound to the resolution in potential. 
For the Au and Si surface this yields a resolution of 60\,mV and 40\,mV respectively.
The resolution for the Au surface is limited by the surface roughness of the contact while the resolution for the Si surface is limited by the slight upwards bend of the potential.
The latter is probably an artifact of the measurement (as the pronounced dip left of the contact is) and not caused by band bending, an effect which one would expect to occur within tens of nanometers from the interface.
The observed bending might be caused by the lateral electric field over the Si area due to the \vb{} applied between the electrodes. 
This field influences the trajectories of the imaging electrons by adding in-plane momentum and therefore, modifying the sample voltage at which the MMT occurs \cite{Jobst-ARRES, Jobst-ARRES-GonBN}.

\begin{figure}[th]
	\includegraphics[width=\columnwidth]{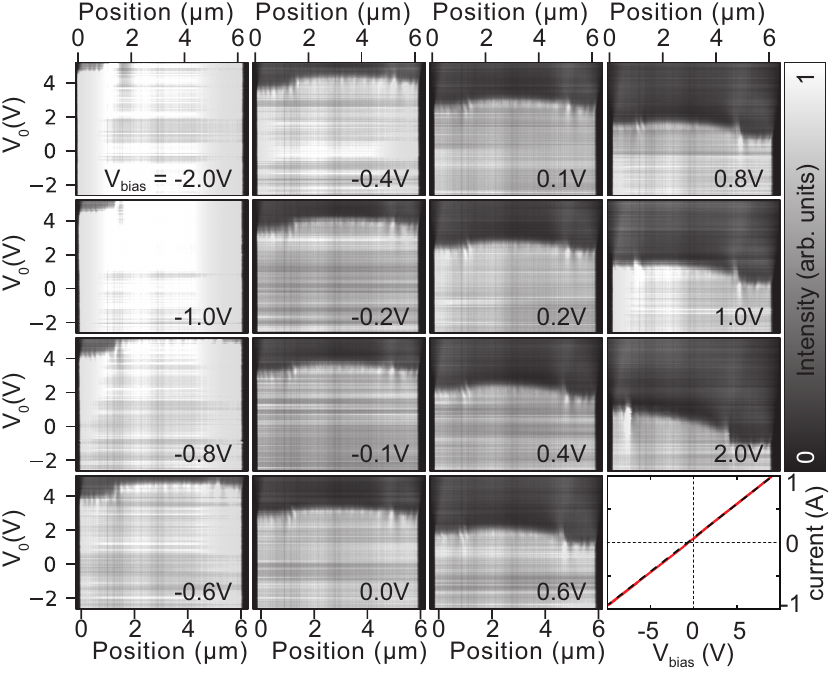}
	\caption{
	Intensity as function of position and \ve{} of the line indicated in Fig. \ref{fig:slices}a.
	For positive bias voltages, there is a voltage drop over the right metal-Si interface, while there is none over the left.
	For negative bias voltages, the situation is reversed.
	}
	\label{fig:vbsweep}
\end{figure}
		
In order to understand the potential step at the metal-Si interface of the source electrode in more detail, we perform M-LEEP measurements as a function of the applied bias \vb.
It is worth noting that such a repeated potentiometry measurement is only feasible because acquiring an M-LEEP data set takes only one minute in contrast to potential measurements with scanning probe techniques.
Figure \ref{fig:vbsweep} shows IV-curves for all pixels on a line across the junction, similar to Fig.\ \ref{fig:slices}c for bias voltages from $\vb = -2$\,V to $\vb = 2$\,V.
For all \vb{} used, potential steps at both electrodes are visible that are composed of a constant part and a part that depends on \vb{}.
The constant part is the workfunction difference between the Au electrodes and the Si(111) surface (similar to Sec.\ \ref{sec:WF}) and can be determined from the case of $\vb = 0$ as $\Delta\Phi_{\text{Au-Si}} \approx 0.35$\,eV in agreement with literature \cite{WF-Si, WF-Au}. 
The \vb -dependent part has a more complicated, non-linear behavior: For small positive \vb, most of the voltage drops over the right metal-Si interface, while the Si surface and the left electrode remain at an unchanged, constant potential. 
Only for $\vb > 0.6$\,V, a clear voltage gradient over the silicon is observed while the steep potential drop over the right metal-Si interface persists.
For negative \vb\, the same behavior can be observed at the left metal-Si interface. 
This non-linear behavior can be explained by Schottky contacts formed between the metallic electrodes and the semiconducting Si. As a consequence, the interface is transparent for the current if it is forward biased causing only a negligible voltage drop at the interface while a sizable fraction of the applied voltage drops over the Schottky contact when it is reverse biased. 
In the full metal-Si and Si-metal device studied, the metal-Si and the Si-metal Schottky junctions at the left and right side, respectively, behave like two diodes with inverted polarities placed in series. 
Consequently, for positive \vb, the right metal-Si interface is reverse biased (large potential step), while the left metal-Si interface is forward biased (small potential step). For negative bias voltages, the situation is reversed.
In other words, the junctions are reverse biased when a positive voltage is applied to the metal contact. This suggests that our Si substrate is slightly p-doped \cite{Laughton2002} due to the cleaning and annealing procedures used.
		
These experiments clearly demonstrate how M-LEEP provides detailed information on local conductance properties. This is particularly important as these local contact effects are invisible in a conventional 2-point resistance measurement. The current versus \vb-measurement shown in the last panel of Fig.\ \ref{fig:vbsweep} is almost perfectly linear, averaging over both Schottky diodes and thus, obscuring the true physics that govern the device performance.

\subsection{Robustness of M-LEEP}
In Sec.\ \ref{sec:WF} and \ref{sec:Schottky}, we have shown that M-LEEP is a versatile tool to determine surface potentials, due to intrinsic workfunction differences or due to externally applied voltages. The advantage of using the steep MMT is that it is universally available for all materials. A notable complication arises, however, if the material studied has a bandgap right at the vacuum level. This bandgap causes low reflectivity over its whole energy range, which leads to an apparent shift of the MMT \cite{Flege-IV, fujikawa-silver}.

We have seen in Fig.\ \ref{fig:potmapgap} that the electrons are easily affected by surface roughness, three-dimensionality of the sample or lateral electric fields due to their low kinetic energy around the MM. This limits the lateral resolution of M-LEEP to values well below the optimal resolution of LEEM \cite{Kennedy-MM-distortions, Kennedy-MM-distortions2}.
Various materials, on the other hand, show other pronounced features in their IV-curves at higher energies where the electron trajectories are less affected by these distortions.  We will demonstrate in the following that those features, instead of the MMT, can be used to determine the local potential, and that this approach leads to better lateral resolution \cite{Kautz-LEEP}.

\begin{figure}[htp]
	\includegraphics[width=\columnwidth]{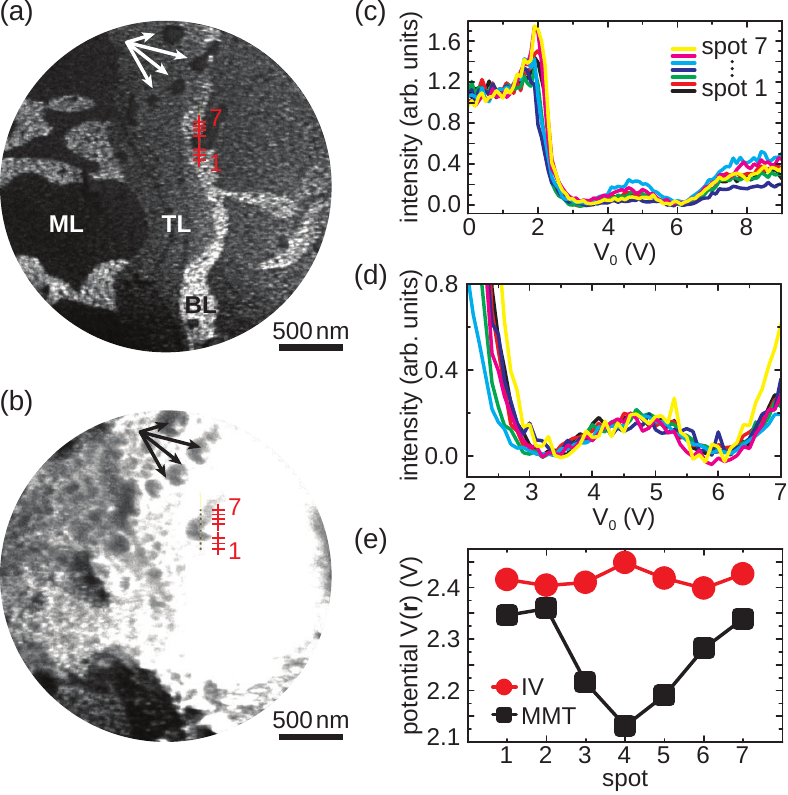}
	\caption{
	(a) LEEM image ($\ve = 4.1$\,V) of monolayer (ML), bilayer (BL) and trilayer (TL) graphene with $\vb = -3$\,V applied from left to right (electrodes are not shown). 
(b) Same area recorded close to MMT ($\ve = 2.4$\,V). The apparent size of imperfections (e.g. arrows) increases close to the MMT. 
(c) IV-curves taken at the points marked in red in a,b exhibit a clear difference in the exact shape of the drop of the MMT. 
(d) The normalized IV-curves from c coincide for higher energies, while the shift in MM is even enhanced. 
(e) The local potential, determined by using the MMT (black squares) or shape of the IV-curve (red circles) as criterion. The former shows strong variations with position although the measurement spots lie on what is to be expected an equipotential line, while the latter exhibits much less variation. The data underlying this figure is also published as Supporting Material of Ref.\ \cite{Kautz-LEEP}.	 
	}
	\label{fig:MMTvsIV}
\end{figure}
				
Figure \ref{fig:MMTvsIV}a shows a LEEM image of an area of monolayer, bilayer and trilayer graphene grown on insulating silicon carbide as described in Ref.\ \cite{emtsev-natmat} with $\vb = -3$\,V applied from left to right. The structure of the sample is clearly visible at the used $\ve = 4.1$\,V, while in Fig.\ \ref{fig:MMTvsIV}b, which is recorded close to the MMT, most of the detail is lost. For example, the apparent size of the imperfections marked with arrows is greatly increased in Fig.\ \ref{fig:MMTvsIV}b compared to a. 
We record IV-curves for points around an imperfection that lie on an equipotential line (Fig.\ \ref{fig:MMTvsIV}c). As expected, the IV-curves look very similar and no large shift is observed in their MMT. However, the exact shape of the MMT varies from spot to spot indicating that problems could occur if the M-LEEP method as described in Sec.\ \ref{sec:potentiometryinleem} is used. 
In fact, if one zooms in on the normalized IV-curves in Fig.\ \ref{fig:MMTvsIV}d, this problem becomes immediately apparent. 
As a consequence, the local potential as determined by M-LEEP exhibits a pronounced lateral variation (black squares in Fig.\ \ref{fig:MMTvsIV}e). This is unexpected as the sampled points lie on an anticipated equipotential line.
If the minima and maxima at higher energy are used as a criterion to calculate the shift of the IV-curves rather than the MMT, no such potential variation is observed (red circles in Fig.\ \ref{fig:MMTvsIV}e) as it is expected from the placement of the measurement points.
This comparison confirms that features of the IV-curve at higher energy, if available, are more robust as a basis for LEEP \cite{Kautz-LEEP}. 
This is particularly important as samples for such transport measurements are often processed by nanofabrication and thus contaminated with residues of lithography resists and other chemicals.

\begin{figure}[th]
	\includegraphics[width=\columnwidth]{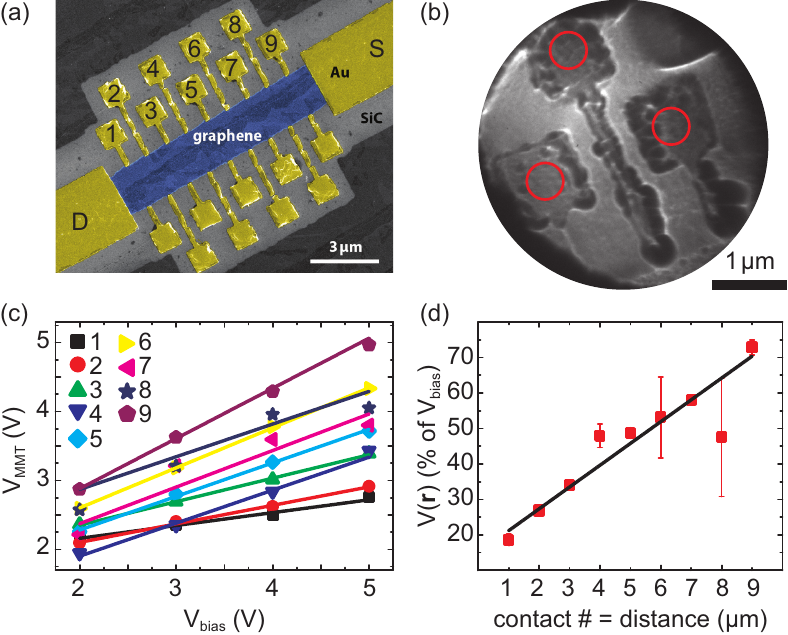}
	\caption{
	(a) Scanning electron micrograph of a graphene Hall bar with Cr/Au electrodes. 
	(b) LEEM image showing some contact pads. IV curves are taken in the marked areas.
	(c) The voltage \vemr{} for which the MMT occurs, is a linearly dependent on the applied bias \vb. Lines are linear fits. 
	(d) The potential determined by M-LEEP as the slope of the fits in a on the center of contact pads shows the expected linear voltage drop. 
	}
	\label{fig:MMTonContacts}
\end{figure}

A way to perform M-LEEP on such processed devices from materials with featureless IV-curves (e.g.\ silicon as seen in Sec.\ \ref{sec:Schottky}) is outlined in Fig.\ \ref{fig:MMTonContacts}. 
A Hall bar with a graphene channel (Fig.\ \ref{fig:MMTonContacts}a) is patterned by electron-beam lithography and subsequent oxygen plasma etching. 
Source and drain contacts as well a multiple voltage probes are defined from Cr/Au (5\,nm/30\,nm) using lithography and evaporation.  
The source and drain contacts are connected to the LEEP sample holder as described in Sec.\ \ref{sec:potentiometryinleem} and a bias voltage \vb{} is applied between them. 
The metallic probes to measure the local voltage drop, on the other hand, are too small to be connected to external measurement equipment but can easily be resolved in LEEM (cf.\ Fig.\ \ref{fig:MMTonContacts}b). 
Performing M-LEEP on these 1\,\textmu m$^2$ big contact pads yields values for the local \vemr{}. These values are plotted as a function of the applied \vb{} for contacts 1 to 9 in Fig.\ \ref{fig:MMTonContacts}c. The expected linear trend following Ohm's law is clearly visible. 
The slope of these linear fits yields the percentage of \vb{} that drops between the according contact and drain. 
In Fig.\ \ref{fig:MMTonContacts}d, these values are plotted over the contact number that is also the distance (in micrometer) from drain. A linear trend is apparent, the slope of which matches the geometry of the Hall bar. Its slight offset is caused by the resistance of the cables, the connections on the chip and the contact resistance between the contacts and graphene, all in series with the channel. 
While this behavior indicates the reliability of M-LEEP, the outlier of contact 8, which is caused by dirt on the contact, also strengthens the point that cleanliness is of utmost importance for M-LEEP measurements.

\section{Conclusions}
Even for simple devices, charge transport can be dominated by local features making it crucial to resolve the potential landscape laterally.
Here we show that LEEM-based potentiometry offers a wide range of opportunities as a versatile tool for that purpose.
We demonstrate that the MMT, available in the IV-curve of any material, allows us to determine the surface potential of virtually any sample (with the small exception of materials that have a bandgap directly at the vacuum level) that can be imaged in LEEM.
Introducing M-LEEP, we study the intrinsic surface potential changes induced by the different workfunctions of the $1\times1$ and the $7\times7$ reconstruction of the Si(111) surface as well as the potential drop induced by applying an external bias over a metal-semiconductor-metal junction. 
We conclude from the latter that a Schottky contact is formed at the metal-Si interface. Remarkably, this local conductance property is invisible in conventional two-point resistance measurements, while we can directly visualize its behavior for various applied voltages using M-LEEP. 
We demonstrate a lateral resolution of 40\,nm and a potential resolution of 40\,mV using M-LEEP.
The resolution here is limited by the fact that the electron energy around the MM is minimal and hence their trajectories are easily deformed by impurities or asperity of the surface. The robustness against such distortions of LEEP can consequently be improved by using features of the IV-curve at higher energy if such features exist. 
We demonstrate that for the example of few layer graphene that exhibits pronounced oscillations in the IV-curve between 0\,eV and 5\,eV. 
We conclude that while M-LEEP cannot reach the lateral resolution of such IV-LEEP, M-LEEP is more versatile in terms of the materials that can be studied since a clear MMT can be observed on nearly all surfaces that can be imaged with LEEM.

\section*{Acknowledgments}
We are grateful to Ruud van Egmond, Martijn Witlox, Raymond Koehler, Leendert Prevo, Marcel Hesselberth, and Daan Boltje for technical assistance, and to Alexander van der Torren, Dani\"el Geelen, Aniket Thete, Jan van Ruitenbeek, and Jan Aarts for useful discussions. This work was supported by the Netherlands Organization for Scientific Research (NWO) via an NWO-Groot grant (\lq ESCHER\rq) and a VIDI grant (680-47-502, SJvdM) and a VENI grant (680-47-447, J.J.) as well as by the FOM foundation via the \lq Physics in 1D\rq\ program.

\section*{References}
\bibliography{ZZ_own-papers-2017_LEEP-Ultramicroscopy}
\bibliographystyle{unsrt}

\end{document}